\documentclass[%
 reprint,
 superscriptaddress,
 amsmath,amssymb,
 aps,
 prx,
]{revtex4-2}

\usepackage[colorlinks=true,
            linkcolor=blue,
            citecolor=blue,
            urlcolor=blue]{hyperref}

\usepackage{graphicx}
\usepackage{dcolumn}
\usepackage{bm}

\begin{document}

\preprint{APS/123-QED}

\title{Supercoiling Drives DNA--Protein Co-Condensation}

\author{Sihan Liu}
 \email{sihanliu@princeton.edu}
 \affiliation{Department of Mechanical and Aerospace Engineering, Princeton University, Princeton, NJ 08544, USA}
 \affiliation{Department of Physics, University of California, San Diego, La Jolla, CA 92093, USA}

\author{Hongbo Zhao}
 \email{hongbo@ucsd.edu}
 \affiliation{Department of Physics, University of California, San Diego, La Jolla, CA 92093, USA}
 \affiliation{Department of Chemistry and Biochemistry, University of California, San Diego, La Jolla, CA 92093, USA}

\author{Andrej Ko\v{s}mrlj}
 \email{andrej@princeton.edu}
 \affiliation{Department of Mechanical and Aerospace Engineering, Princeton University, Princeton, NJ 08544, USA}
 \affiliation{Princeton Materials Institute, Princeton University, Princeton, NJ 08544, USA}

\date{\today}

\begin{abstract}

The co-condensation of DNA and proteins plays an important role in genome compaction, including heterochromatin formation and chromosome segregation. Meanwhile, transcription and replication continuously generate DNA supercoiling, yet how the resulting torsional stress affects co-condensation remains poorly understood. Here, by coupling DNA elasticity to a protein concentration field, we show that supercoiling acts as a topological catalyst for co-condensation through a mutually reinforcing interplay with protein wetting. Specifically, protein wetting facilitates twist-to-writhe conversion, generating highly writhed DNA conformations that in turn promote capillary collapse into a protein-rich droplet. In the absence of supercoiling, DNA remains kinetically trapped in metastable extended states within our model, whereas sufficiently strong supercoiling removes the nucleation barrier and enables spontaneous co-condensation. These results establish topology as a distinct control parameter for biomolecular condensation and provide a mechanical framework for how active genomic processes can dynamically regulate large-scale genome organization.

\end{abstract}

\maketitle

{\it Introduction}.
The spatial organization and compaction of DNA are central to genome function across all domains of life. In cells, DNA can adopt a variety of condensed architectures, ranging from heterochromatin in eukaryotes \cite{grewal2007heterochromatin, strom2017phase, strom2019liquid, sabari2020biomolecular, lee2022mechanobiology, xia2025chromatin, zhao2025condensate} to compact bacterial chromosomes during segregation \cite{wang2013organization, badrinarayanan2015bacterial, tivsma2023dynamic}. In many such systems, proteins associate with DNA and provide attractive interactions that thermodynamically favor protein-assisted DNA condensation \cite{brackley2013nonspecific, broedersz2014condensation, larson2017liquid, ryu2021bridging, renger2022co, connolley2023partition, strom2024interplay, chappidi2024parp1, haugk2024single, ahlawat2026measuring}. Yet thermodynamic favorability does not necessarily imply rapid compaction: long DNA molecules can remain kinetically trapped in extended conformations, separated from condensed states by substantial nucleation barriers \cite{ueda1996phase, quail2021force, shahu2024bridging, takaki2025active}. Overcoming these barriers can therefore be essential for DNA compaction to occur on biologically relevant timescales.

Cells can accelerate genome reorganization through active processes that drive DNA away from kinetically trapped configurations. A well-studied example is loop extrusion, in which motor proteins reel in distal DNA segments and organize them into loops, thereby promoting rapid compaction \cite{kim2022condensin, takaki2025active}. Transcription and replication provide another ubiquitous source of nonequilibrium mechanical input: as polymerases move along the DNA double helix, they generate torsional stress that leads to DNA supercoiling \cite{vologodskii1994conformational, koster2010cellular, junier2023dna}.

DNA supercoiling can reshape DNA into highly coiled, interwound structures such as plectonemes, partially relaxing the torsional elastic energy stored in the double helix \cite{marko1994fluctuations, marko1995statistical}.
By bringing distant DNA segments into close spatial proximity, these highly coiled conformations may strongly alter the pathway by which DNA and proteins co-condense. Indeed, recent single-molecule experiments have shown that supercoiling can accelerate DNA--protein co-condensation and lower the critical protein concentration required for condensation \cite{martin2024dna, shahu2024bridging}. Yet an {\it in vivo} study reported only weak sensitivity of condensation to DNA supercoiling \cite{alaoui2025vivo}. These contrasting observations suggest that the effect of supercoiling is strongly regime-dependent, and that the physical mechanism coupling DNA topology, elasticity, and protein-mediated condensation remains unresolved.

\begin{figure}[!b]
	\centering
	\includegraphics[width=0.8\linewidth]{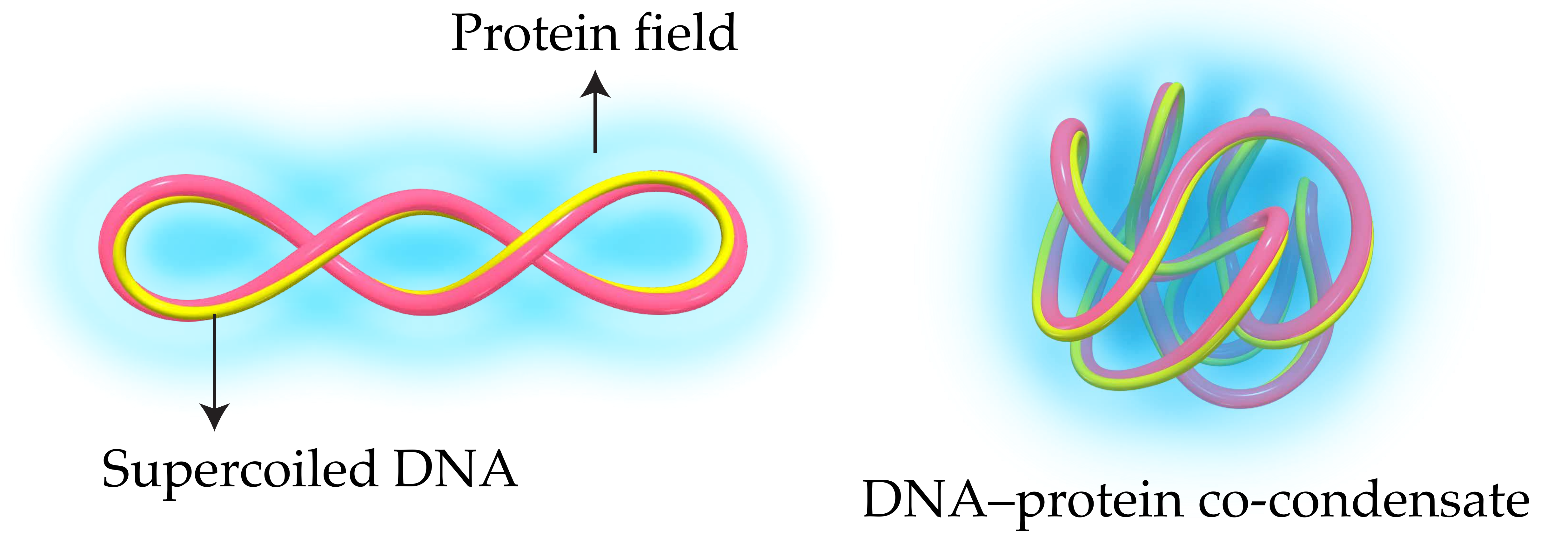}
	\caption{
    Schematics of DNA supercoiling and DNA--protein co-condensation. Pink and yellow curves denote the DNA centerline and its associated material frame, respectively. Blue shading represents the protein concentration field, with proteins accumulating around the DNA contour through wetting. (Left) Plectonemic DNA wetted by proteins at excess linking number $\Delta{\rm Lk}=4$. (Right) A fully collapsed DNA--protein co-condensate, in which DNA is encapsulated within a protein-rich droplet ($\Delta{\rm Lk}=9$).
    }
	\label{model}
\end{figure}

{\it Continuum Model}.
To investigate how DNA supercoiling influences protein-assisted condensation, we develop a theoretical framework that couples DNA elasticity to a diffusible protein concentration field (Fig.~\ref{model}). DNA is modeled as an elastic closed loop of contour length $L$, characterized by an excess linking number $\Delta{\rm Lk}$, which measures the change in the total winding of the DNA double helix relative to a torsionally relaxed reference state with a planar circular centerline. For a closed DNA loop, $\Delta{\rm Lk}$ is a topological invariant conserved under smooth deformations that do not involve strand crossings \cite{crick1976linking}. It is partitioned into excess twist $\Delta{\rm Tw}$, describing over- or underwinding of the double helix relative to the torsionally relaxed state, and writhe ${\rm Wr}$, describing the three-dimensional coiling of the DNA centerline, such that $\Delta{\rm Lk}=\Delta{\rm Tw}+{\rm Wr}$ \cite{crick1976linking}. The linking density $\Delta{\rm Lk}/L$ serves as the control parameter for the degree of supercoiling in our model. It can be related to the conventional dimensionless supercoiling density through $\sigma=h\Delta{\rm Lk}/L$ \cite{vologodskii1994conformational}, where $h$ is the helical pitch of the relaxed DNA double helix. At sufficiently large $|\Delta{\rm Lk}|$, DNA can reduce its torsional elastic energy by converting excess twist into writhe, leading to buckling and the formation of plectonemic structures (Fig.~\ref{model}, left).

\begin{figure}[!t]
	\centering
	\includegraphics[width=1\linewidth]{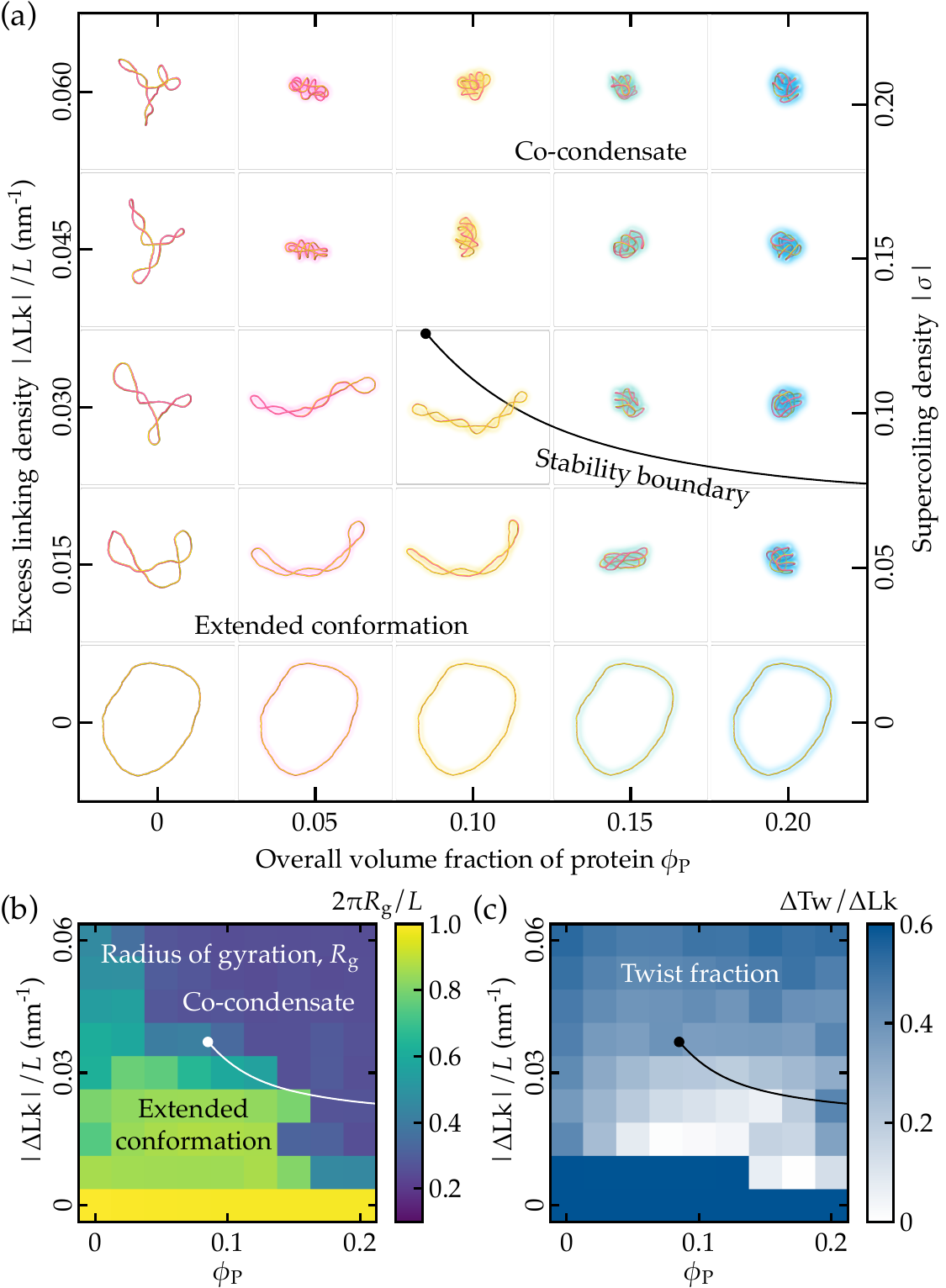}
	\caption{
    Phase diagram of supercoiled DNA conformations and protein-mediated co-condensation. (a) DNA conformations as a function of the mean protein volume fraction $\phi_{\rm P}$ and excess linking density $\Delta{\rm Lk}/L$. The corresponding supercoiling density is $\sigma=h\Delta{\rm Lk}/L$, where $h\approx3.4\,\mathrm{nm}$ is the helical pitch of relaxed DNA. All simulations are initialized from a circular DNA loop of contour length $L=200\,\mathrm{nm}$. Representative dynamical evolutions are shown in Supplemental Movie~1. (b, c) Heat maps over the same parameter space showing (b)~the time-averaged DNA radius of gyration $R_{\rm g}$ and (c)~the time-averaged fraction of the excess linking number accommodated as twist, $\Delta{\rm Tw}/\Delta{\rm Lk}$. Near $\Delta{\rm Lk}=0$, this ratio can exceed the displayed color range because thermal fluctuations produce finite $\Delta{\rm Tw}$ while $\Delta{\rm Lk}$ is small; at $\Delta{\rm Lk}=0$, the ratio is formally undefined. The stability boundary shown here is obtained from the analytical theory introduced in Fig.~\ref{minimum_model}.
    }
	\label{phase_diagram}
\end{figure}

To model DNA--protein interactions, we represent DNA using a continuous volume-fraction field localized along its contour and couple it to protein and solvent volume-fraction fields through a Flory--Huggins free-energy functional with interfacial gradient terms \cite{flory1942thermodynamics, huggins1941solutions, cahn1958free, liu2025predicting}. We consider a minimal regime in which DNA interacts neutrally and symmetrically with protein and solvent, while protein and solvent repel each other. We restrict our analysis to mean protein volume fractions $\phi_{\rm P}\le 0.2$, below the saturation concentration of the DNA-free protein--solvent mixture, such that proteins do not spontaneously phase-separate in the bulk. Nevertheless, the unfavorable protein--solvent interaction drives local protein enrichment along the DNA contour, producing a wetting layer around DNA (Fig.~\ref{model}). When distant DNA segments approach one another, their wetting layers overlap, generating an effective capillary attraction that promotes further compaction.
The coupled system evolves through overdamped relaxation of the DNA conformation and Cahn--Hilliard dynamics of the protein concentration field; full equations and model parameters are provided in Appendix~A of the Supplemental Material.

{\it Phase Diagram of Supercoiled DNA--Protein Systems}.
To elucidate the interplay between protein wetting and DNA supercoiling, we systematically vary the mean protein volume fraction $\phi_{\rm P}$ and the excess linking density $\Delta{\rm Lk}/L$ (Fig.~\ref{phase_diagram}). The corresponding supercoiling densities span $|\sigma|\lesssim0.2$, with $|\sigma|\lesssim0.1$ encompassing values commonly observed in bacterial DNA; for example, the \textit{E.~coli} chromosome is typically maintained at $|\sigma|\sim0.05$--$0.07$ \cite{higgins2016species, visser2022psoralen}. In the absence of protein ($\phi_{\rm P}=0$), DNA with zero excess linking number adopts a highly expanded, fluctuating conformation, whereas increasing $|\Delta{\rm Lk}|$ drives buckling into increasingly writhed plectonemic conformations (Fig.~\ref{phase_diagram}a). At nonzero protein concentrations, proteins preferentially wet the DNA contour and generate an effective capillary attraction between nearby DNA segments. This attraction promotes DNA compaction, as reflected by the decrease in the radius of gyration $R_{\rm g}$ with increasing $\phi_{\rm P}$ (Fig.~\ref{phase_diagram}a, b). Importantly, protein wetting also promotes twist-to-writhe conversion within the extended plectonemic regime:  increasing $\phi_{\rm P}$ substantially reduces the twist fraction $\Delta{\rm Tw}/\Delta{\rm Lk}$ relative to the corresponding protein-free state (Fig.~\ref{phase_diagram}c). Capillary attraction draws the intertwined DNA segments closer together, reducing the plectoneme radius and favoring conformations with greater writhe. Because $\Delta{\rm Lk}=\Delta{\rm Tw}+{\rm Wr}$ is conserved, the increase in writhe reduces the excess twist and releases torsional elastic energy. Protein wetting therefore facilitates the conversion of torsional deformation into conformational coiling. At sufficiently high supercoiling and protein volume fraction, this cooperative effect culminates in complete DNA collapse within a protein-rich droplet, forming a DNA--protein co-condensate (Fig.~\ref{phase_diagram}a; Fig.~\ref{model}, right). Within the parameter range considered here, neither supercoiling nor protein wetting is sufficient to induce complete collapse by itself, as illustrated by the leftmost column and bottom row of Fig.~\ref{phase_diagram}a. The resulting phase diagram thus reveals a cooperative interplay between DNA topology and protein wetting in driving DNA--protein co-condensation.

\begin{figure}[!t]
	\centering
	\includegraphics[width=1\linewidth]{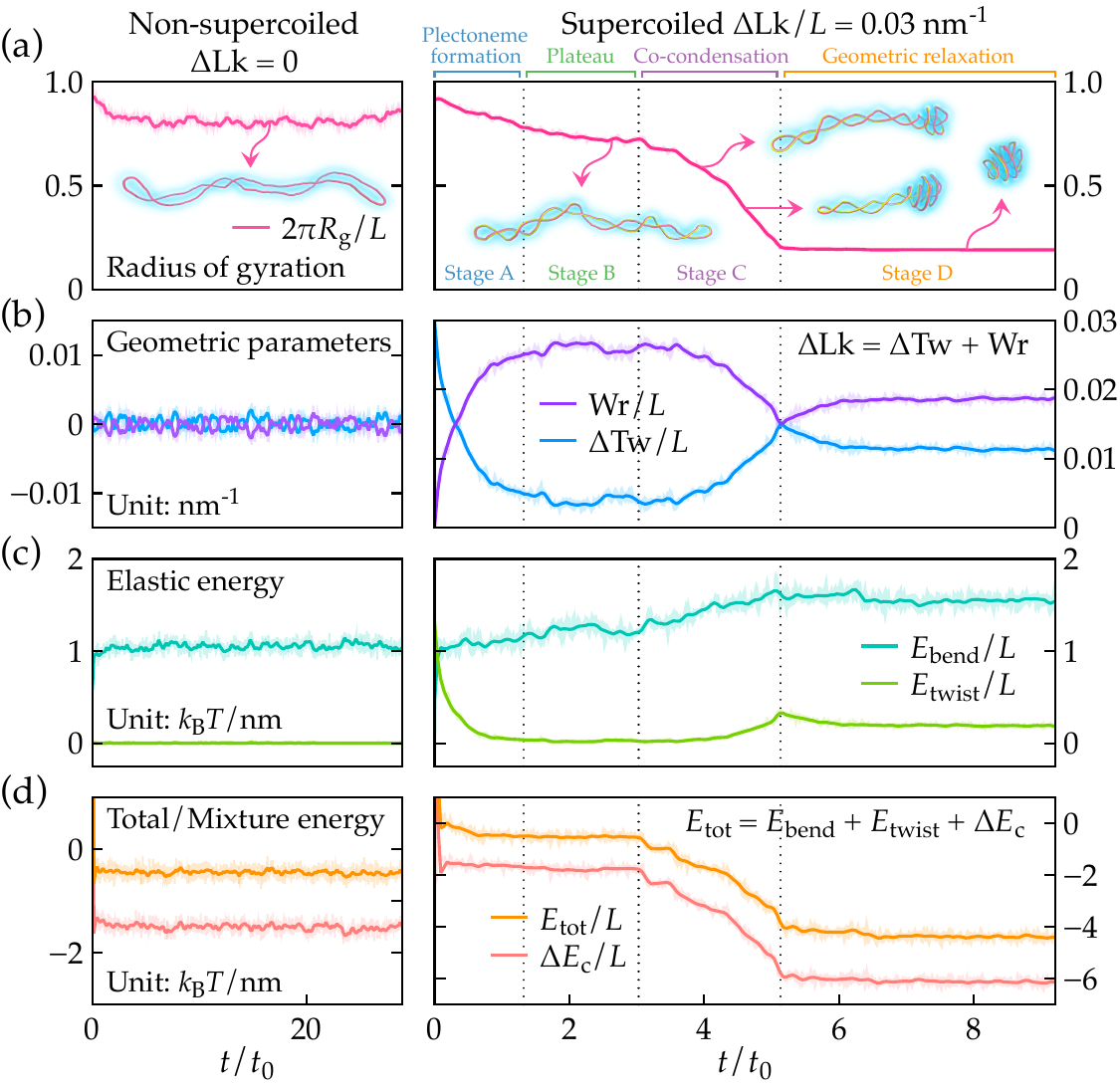}
	\caption{
    Kinetics of DNA--protein systems without (left) and with (right) supercoiling. The rows show the time evolution of (a) the normalized DNA radius of gyration, $2\pi R_{\rm g}/L$; (b) the writhe and excess twist densities, ${\rm Wr}/L$ and $\Delta{\rm Tw}/L$; (c) the bending and twisting energy densities, $E_{\rm bend}/L$ and $E_{\rm twist}/L$; and (d) the total and chemical interaction energy densities, $E_{\rm tot}/L$ and $\Delta E_{\rm c}/L$. Here, $\Delta E_{\rm c}$ is measured relative to a straight, non-self-contacting DNA conformation equilibrated in the same protein solution. The left and right panels correspond to excess linking densities $\Delta{\rm Lk}/L=0$ and $0.03\,\mathrm{nm}^{-1}$, respectively, at a fixed mean protein volume fraction $\phi_{\rm P}=0.2$. Both simulations are initialized from the same hairpin-shaped DNA conformation, consisting of two straight, parallel segments connected at their ends, with contour length $L=300\,\mathrm{nm}$. The characteristic time $t_0$ is the longest Rouse relaxation time of the corresponding ideal ring polymer \cite{doi1988theory, tsolou2010melt} (Supplemental Material, Appendix~A). Insets show representative simulation snapshots at the indicated times. Full dynamical evolutions are shown in Supplemental Movies~2 and~3.
    }
	\label{time_evolution}
\end{figure}

{\it Kinetics of DNA--Protein Co-Condensation}.
The phase diagram in Fig.~\ref{phase_diagram} already points to a strong kinetic constraint: at $\phi_{\rm P}=0.2$, non-supercoiled DNA initialized from a circular conformation remains extended despite protein wetting, even though the fully co-condensed state is thermodynamically favorable (Supplemental Material, Appendix~B). This indicates that co-condensation is hindered by a kinetic barrier rather than by the absence of a thermodynamic driving force. To probe this barrier more directly, we examine the condensation kinetics using a closed hairpin initial geometry consisting of two parallel segments connected at their ends (Fig.~\ref{time_evolution}). This choice removes an additional geometric bottleneck present for an initially circular, non-supercoiled DNA: in the parameter regime studied here, wetting-induced capillary attraction is too weak to bend the circular DNA directly into a collapsed state, so nonadjacent DNA segments must first be brought into close proximity by a fluctuation before capillary attraction can stabilize and amplify their contact. The hairpin initial condition bypasses this encounter-limited delay by placing two DNA segments in close proximity from the outset. Its resemblance to the closely spaced DNA segments in a plectoneme also allows us to ask whether such pre-existing proximity is sufficient for non-supercoiled DNA to condense, and how supercoiling alters the ensuing dynamics.

Figure~\ref{time_evolution} reveals that supercoiling dramatically accelerates DNA--protein co-condensation. The supercoiled system collapses within $t\approx5\,t_0$, whereas non-supercoiled DNA remains in a fluctuating hairpin configuration over an observation window exceeding $25\,t_0$.
Here, $t_0$ denotes the longest Rouse relaxation time of the corresponding ideal ring polymer \cite{doi1988theory, tsolou2010melt} (Supplemental Material, Appendix~A).
The evolution of the supercoiled system can be divided into four dynamical stages (Fig.~\ref{time_evolution}). During stage~A ($t\lesssim\,t_0$), excess twist is rapidly converted into writhe through plectoneme formation (Fig.~\ref{time_evolution}b), producing a sharp decrease in the twisting energy $E_{\rm twist}$ (Fig.~\ref{time_evolution}c). The accompanying increase in bending energy $E_{\rm bend}$, driven by the more strongly coiled conformation, is more than offset by the reduction in twisting energy. Meanwhile, the chemical interaction energy $\Delta E_{\rm c}$ drops sharply at early times (Fig.~\ref{time_evolution}d) as the protein concentration field relaxes around the initial hairpin, establishing wetting-mediated interactions between nearby DNA segments. Following this initial relaxation, $\Delta E_{\rm c}$ changes little during plectoneme formation, indicating that the hairpin-to-plectoneme rearrangement is driven primarily by the release of DNA torsional elastic energy. The resulting plectoneme then fluctuates in a metastable state (stage~B) until a compact domain nucleates at the plectoneme tip at $t\approx3\,t_0$. During stage~C, this domain progressively absorbs the remaining plectoneme, while writhe decreases and excess twist reaccumulates. In contrast to stage~A, the decrease in chemical interaction energy now provides the dominant reduction in total energy $E_{\rm tot}$, outweighing the accompanying increases in bending and twisting energies and driving co-condensation. Following complete collapse, DNA continues to reorganize within the protein-rich droplet (stage~D), slightly reducing its elastic energy as the system approaches equilibrium. Overall, the total energy decreases throughout the process apart from small thermal fluctuations (Fig.~\ref{time_evolution}d). The contrast with the non-supercoiled system is therefore kinetic rather than thermodynamic: although the fully co-condensed state remains thermodynamically favorable at zero supercoiling (Supplemental Material, Appendix~B), the non-supercoiled hairpin remains metastable over the entire observation window. This long-lived metastability reveals a substantial nucleation barrier to co-condensation in the absence of supercoiling.

\begin{figure}[!t]
	\centering
	\includegraphics[width=1\linewidth]{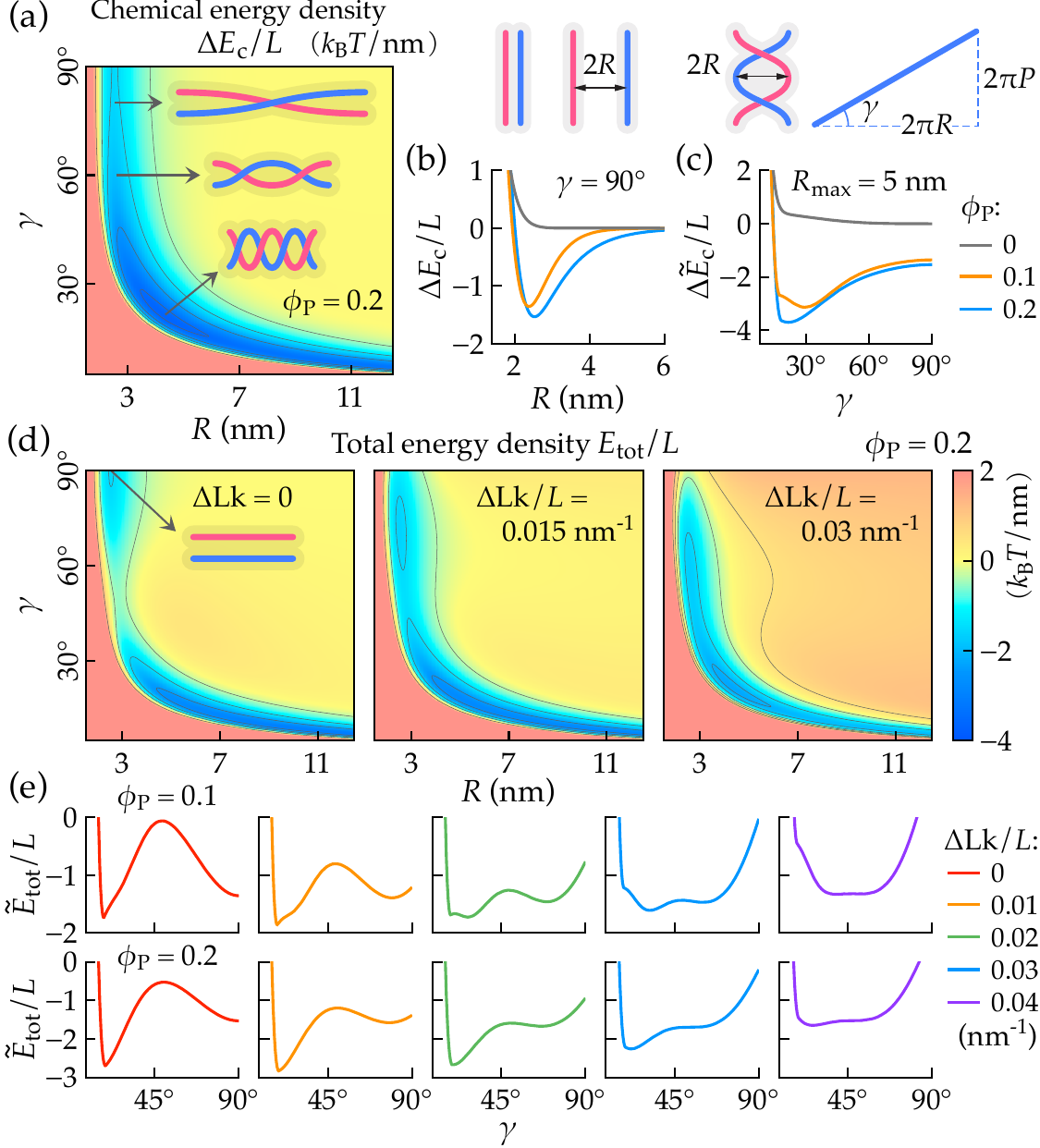}
	\caption{
    Energy landscapes of the reduced model for DNA--protein co-condensation. DNA is constrained to a uniform periodic plectonemic geometry characterized by radius $R$ and helix angle $\gamma$.
    (a) Chemical interaction energy density $\Delta E_{\rm c}/L$ as a function of $R$ and $\gamma$ at far-field protein volume fraction $\phi_{\rm P}=0.2$. The energy is measured relative to two straight, parallel DNA segments at infinite separation, $\Delta E_{\rm c}(R,\gamma)/L=[E_{\rm c}(R,\gamma)-E_{\rm c}(R\to\infty,\gamma=\pi/2)]/L$.
    (b) Chemical interaction energy density for two straight, parallel DNA segments separated by a distance $2R$, shown for different $\phi_{\rm P}$.
    (c) Minimum chemical interaction energy density as a function of helix angle, $\Delta\widetilde E_{\rm c}(\gamma)/L\equiv\min_{R\le R_{\rm max}}\{\Delta E_{\rm c}(R,\gamma)/L\}$, with $R_{\rm max}=5\,\mathrm{nm}$, for different $\phi_{\rm P}$.
    (d) Total energy density $E_{\rm tot}/L$ as a function of $R$ and $\gamma$ at different excess linking densities $\Delta{\rm Lk}/L$, where $E_{\rm tot}=\Delta E_{\rm c}+E_{\rm bend}+E_{\rm twist}$ [Eqs.~(\ref{Eq:Eel})].
    (e) Minimum total energy density as a function of $\gamma$, $\widetilde E_{\rm tot}(\gamma)/L\equiv\min_{R\le R_{\rm max}}\{E_{\rm tot}(R,\gamma)/L\}$, at $\phi_{\rm P}=0.1$ (top) and $0.2$ (bottom).
    }
	\label{minimum_model}
\end{figure}

{\it Energy Landscape of DNA--Protein Co-Condensation}.
To rationalize how supercoiling modifies the nucleation barrier to co-condensation, we introduce a minimal analytical model that resolves the competition between capillary and elastic energies. DNA is constrained to a uniform, infinitely long plectonemic geometry parameterized by its radius $R$ and helix angle $\gamma=\tan^{-1}(P/R)$, where $2\pi P$
is the axial pitch of the plectoneme
(Fig.~\ref{minimum_model}). For each $(R,\gamma)$ at a prescribed far-field protein volume fraction $\phi_{\rm P}$, we equilibrate the protein concentration field and define the chemical interaction energy per unit DNA length relative to two straight, parallel DNA segments at infinite separation as $\Delta E_{\rm c}(R,\gamma)/L=[E_{\rm c}(R,\gamma)-E_{\rm c}(R\to\infty,\gamma=\pi/2)]/L$. At $\phi_{\rm P}=0.2$, the resulting energy landscape (Fig.~\ref{minimum_model}a) shows that along the parallel branch ($\gamma=\pi/2$), capillary attraction favors a finite separation, with an energy minimum near $R\approx2.5\,\mathrm{nm}$ corresponding to a hairpin-like configuration. Decreasing $\gamma$ further lowers $\Delta E_{\rm c}/L$, favoring progressively more compact plectonemic geometries and ultimately a global minimum at small $\gamma$. Within this constrained description, the small-$\gamma$ minimum serves as a reduced representation of the co-condensed state. 
Figure~\ref{minimum_model}b shows $\Delta E_{\rm c}/L$ as a function of the separation between parallel DNA segments ($\gamma=\pi/2$), while Fig.~\ref{minimum_model}c shows the minimized energy $\Delta\widetilde E_{\rm c}(\gamma)/L\equiv\min_{R\le R_{\rm max}}\{\Delta E_{\rm c}(R,\gamma)/L\}$ as a function of $\gamma$, with $R_{\rm max}=5\,\mathrm{nm}$ restricting the minimization to the physically relevant plectonemic branch. Together, these results show that higher protein concentrations strengthen the capillary attraction and more strongly favor tightly coiled plectonemic conformations.

For a uniform plectoneme, the elastic energy consists of bending and twisting contributions \cite{marko1994fluctuations, marko1995statistical},
\begin{subequations} \label{Eq:Eel}
\begin{align}
    \frac{E_{\rm bend}}{k_{\rm B}TL}
    &=
    \frac{\ell_{\rm pb}\cos^4\gamma}{2R^2},
    \\
    \frac{E_{\rm twist}}{k_{\rm B}TL}
    &=
    \frac{\ell_{\rm pt}}{2}
    \left(
    \frac{2\pi\Delta{\rm Lk}}{L}
    - \frac{\sin\gamma\cos\gamma}{R}
    \right)^2,
\end{align}
\end{subequations}
where $\ell_{\rm pb}\approx50\,\mathrm{nm}$ \cite{bouchiat1999estimating} and $\ell_{\rm pt}\approx75\,\mathrm{nm}$ \cite{vologodskii1994conformational} are the bending and twisting persistence lengths, respectively. For comparison, if the excess linking number is initially stored entirely as twist, the twisting energy density can be written as
$E_{\rm twist}/L=(1/2)k_{\rm B}T\ell_{\rm pt}(2\pi\sigma/h)^2$.
For biologically relevant supercoiling densities, $|\sigma|\sim0.05$--$0.1$, this gives $E_{\rm twist}/L\sim0.3$--$1.3\,k_{\rm B}T/\mathrm{nm}$, comparable to the wetting-induced capillary energy density of order $k_{\rm B}T/\mathrm{nm}$ between nearby DNA segments in Fig.~\ref{minimum_model}a--c. 
Figure~\ref{minimum_model}d shows the resulting total energy landscapes at $\phi_{\rm P}=0.2$. At zero supercoiling, the landscape contains two minima: a metastable hairpin-like state at $\gamma=90^\circ$ and $R\approx2.5\,\mathrm{nm}$, and a lower-energy compact state at small~$\gamma$, separated by a finite energy barrier. By contrast, at $\Delta{\rm Lk}/L=0.03\,\mathrm{nm}^{-1}$, the barrier disappears and only the compact small-$\gamma$ minimum remains. Thus, within the reduced energy landscape, supercoiling can eliminate the nucleation barrier and render the transition toward co-condensation barrierless, explaining why non-supercoiled DNA remains kinetically trapped whereas supercoiled DNA rapidly collapses in the full simulations (Figs.~\ref{phase_diagram} and~\ref{time_evolution}).

Figure~\ref{minimum_model}e shows the minimum total energy density as a function of $\gamma$ for varying excess linking densities at $\phi_{\rm P}=0.1$ and~$0.2$, including a bistable regime with competing extended and compact minima.
We use this reduced theory to determine the stability boundary of the extended state, shown by the solid line in Fig.~\ref{phase_diagram}, along which the nucleation barrier vanishes. Above and to the right of this boundary, the extended minimum disappears and collapse becomes barrierless within the reduced model, consistent with the spontaneous co-condensation observed in the full simulations. On the other side, a finite barrier persists and can kinetically trap DNA in an extended state over the simulation timescale. The predicted stability boundary terminates at a critical point of the reduced energy landscape (dot in Fig.~\ref{phase_diagram}), where the bistable region ends and only a single local minimum remains. Because the analytical model is restricted to uniform plectonemic geometries, it does not capture the final spherical morphology observed in the unconstrained simulations: when the preferred helix angle $\gamma$ becomes sufficiently small, DNA can further collapse into a compact DNA--protein co-condensate (Figs.~\ref{phase_diagram} and~\ref{time_evolution}; Appendix~C of the Supplemental Material).

{\it Discussion}. Through combined simulations and theoretical analysis, we demonstrate that DNA supercoiling provides a powerful mechanism for promoting DNA--protein co-condensation. While condensation is often attributed to sufficiently strong protein--DNA interactions, supercoiling offers a distinct physical route: it lowers the nucleation barrier by biasing DNA toward writhed conformations, thereby reshaping the condensation pathway. In this process, protein wetting facilitates twist-to-writhe conversion, and the resulting writhed geometry further enhances wetting-induced capillary attraction, revealing a mutually reinforcing coupling between topology and wetting.
For the biologically relevant supercoiling range $|\sigma|\sim0.05$--$0.1$ considered here, the characteristic torsional energy density is $\sim0.3$--$1.3\,k_{\rm B}T/\mathrm{nm}$.
In our model, the wetting-induced capillary energy density is $\sim1\,k_{\rm B}T/\mathrm{nm}$ for nearby DNA segments and reaches $\sim6\,k_{\rm B}T/\mathrm{nm}$ upon full condensation, corresponding to $\sim0.3$ and $2\,k_{\rm B}T/\mathrm{bp}$, respectively, or energy-per-length scales of roughly $4$ and $25\,\mathrm{pN}$. These scales are consistent in order of magnitude with energetic estimates from single-molecule studies of protein-mediated DNA compaction, including $\sim0.8\,k_{\rm B}T/\mathrm{bp}$ for HIV-1 nucleocapsid--DNA self-attraction \cite{gien2022hiv} and a $\sim1\,k_{\rm B}T/\mathrm{bp}$ mechanical disassembly barrier for HP1$\alpha$--DNA condensates \cite{keenen2021hp1}. Related protein--DNA condensates also exhibit measured mechanical forces ranging from $\sim0.2$ to $35\,\mathrm{pN}$ \cite{quail2021force, bell2021single, nguyen2022chromatin, ahlawat2026measuring}. These comparisons indicate that the interaction strengths considered here lie within experimentally relevant scales. Crucially, the comparable magnitudes of DNA torsional elasticity and protein-mediated capillary attraction place the system in a regime where supercoiling can substantially reshape the nucleation barrier and thereby promote co-condensation.

Our framework offers a physical basis for reconciling seemingly disparate experimental observations. The strong enhancement of condensation by supercoiling observed {\it in vitro} \cite{martin2024dna, shahu2024bridging} corresponds naturally to a regime in which protein-mediated attraction alone is insufficient to overcome the nucleation barrier, making supercoiling a decisive factor in enabling collapse. Conversely, the weak sensitivity to supercoiling reported {\it in vivo} \cite{alaoui2025vivo} could reflect an affinity-dominated regime in which sufficiently strong DNA--protein interactions largely eliminate this barrier. Condensation then occurs readily even at zero supercoiling, rendering the effect of supercoiling unnecessary.

Because transcription continuously generates local DNA supercoiling, our results suggest that supercoiling may facilitate the formation of transcriptional condensates at active loci \cite{sabari2018coactivator, cho2018mediator, guo2019pol}. More generally, supercoiling may provide a dynamically tunable control parameter for genome compaction. Supercoiling can be locally generated, propagated, and relaxed \cite{vologodskii1994conformational, koster2010cellular, junier2023dna}, enabling spatiotemporal regulation without changing protein composition. Our theory further suggests hysteresis: once formed, the co-condensate remains thermodynamically stable even after supercoiling is removed. Transient supercoiling can therefore act as a kinetic trigger for long-lived DNA compaction.

More broadly, our findings establish a general principle: coupling polymer topology to wetting-induced capillary interactions can reshape condensation pathways by removing kinetic barriers. This mechanism may extend beyond DNA to other twisted semiflexible polymers and filaments interacting with wetting media, such as protein-rich condensates that wet cytoskeletal filaments and liquid droplets that drive elastocapillary bundling of synthetic fibers.

{\it Acknowledgement}. We thank Ned Wingreen for valuable discussions. This work was supported by the National Science Foundation through Princeton University’s Materials Research Science and Engineering Center (DMR-2011750, DMR-2614599).

\bibliography{Supercoiling}

\end{document}


\preprint{APS/123-QED}

\title{Supplemental Material: Supercoiling Drives DNA–Protein Co-Condensation}

\author{Sihan Liu}
 \email{sihanliu@princeton.edu}
 \affiliation{Department of Mechanical and Aerospace Engineering, Princeton University, Princeton, NJ 08544, USA}
 \affiliation{Department of Physics, University of California, San Diego, La Jolla, CA 92093, USA}

\author{Hongbo Zhao}
 \email{hongbo@ucsd.edu}
 \affiliation{Department of Physics, University of California, San Diego, La Jolla, CA 92093, USA}
 \affiliation{Department of Chemistry and Biochemistry, University of California, San Diego, La Jolla, CA 92093, USA}

\author{Andrej Ko\v{s}mrlj}
 \email{andrej@princeton.edu}
 \affiliation{Department of Mechanical and Aerospace Engineering, Princeton University, Princeton, NJ 08544, USA}
 \affiliation{Princeton Materials Institute, Princeton University, Princeton, NJ 08544, USA}

\date{\today}

\maketitle

\section{Elastocapillary Model}

In our model, DNA is represented as a closed semiflexible chain with bending and twisting elasticity. Focusing on these contributions, we write the elastic free energy as \cite{marko1994fluctuations, marko1995statistical}
\begin{align}
    \frac{E_{\rm el}}{k_{\rm B}T} =
    \frac{\ell_{\rm pb}}{2}
    \int {\rm d}s
    \left(
    \frac{\partial^2 \mathbf r}{\partial s^2}
    \right)^2
    +
    \frac{\ell_{\rm pt}\left(2\pi\Delta{\rm Tw}\right)^2}{2L}.
\end{align}
Here, $k_{\rm B}$ is the Boltzmann constant, $T$ is the temperature, and $\ell_{\rm pb}\approx50\,\mathrm{nm}$ \cite{bouchiat1999estimating} and $\ell_{\rm pt}\approx75\,\mathrm{nm}$ \cite{vologodskii1994conformational} are the bending and twisting persistence lengths, respectively. The variable $s$ is the contour coordinate along the DNA, $\mathbf r(s)$ is the three-dimensional position of the DNA centerline, and $L$ is the total contour length. The twisting contribution assumes that the excess twist is uniformly distributed along the DNA contour. The total excess twist satisfies $\Delta{\rm Tw}=\Delta{\rm Lk}-{\rm Wr}$, with the writhe of a closed loop given by \cite{fuller1971writhing, white1968self}
\begin{align}
    {\rm Wr} = \frac{1}{4\pi}  \iint 
    {\rm d}s_1  {\rm d}s_2 
    \frac{\left[ \partial_{s_1} \mathbf r(s_1) \times \partial_{s_2} \mathbf r(s_2) \right] \cdot \left[\mathbf r(s_1) - \mathbf r(s_2)\right]}{|\mathbf r(s_1) - \mathbf r(s_2)|^3}.
\end{align}
Beyond the leading quadratic elastic terms retained here, double-stranded DNA exhibits twist--stretch coupling, whereby torsional deformation can induce axial extension and vice versa \cite{gore2006dna}. In our model, the DNA contour is nearly inextensible, and no external stretching force is applied. We therefore neglect twist--stretch coupling and higher-order torsional elasticity \cite{bryant2003structural} to isolate the interplay between DNA supercoiling and protein-mediated capillary attraction. The resulting minimal model treats positive and negative supercoiling symmetrically.


The chemical interactions between DNA and the surrounding protein--solvent mixture are described by a Flory--Huggins free energy of mixing \cite{flory1942thermodynamics, huggins1941solutions}, together with a Cahn--Hilliard gradient contribution \cite{cahn1958free},
\begin{align} \label{Eq:E_c}
    \frac{E_{\rm c}}{\rho_0k_{\rm B}T}
    = \int {\rm d}^3\mathbf x
    \left[
    \phi_{\rm P}\ln\phi_{\rm P}
    + \phi_{\rm S}\ln\phi_{\rm S}
    + \chi_{\rm PS}\phi_{\rm P}\phi_{\rm S}
    - \lambda^2 \chi_{\rm PS}
    \nabla\phi_{\rm P} \cdot \nabla\phi_{\rm S}
    \right].
\end{align}
Here, $\rho_0=50\,\mathrm{nm}^{-3}$ is the number density of microscopic lattice sites, $\mathbf x$ is the three-dimensional spatial coordinate, and $\phi_{\rm P}$ and $\phi_{\rm S}$ are the volume-fraction fields of protein and solvent, respectively. The parameter $\lambda=1\,\mathrm{nm}$ sets the characteristic interfacial length scale, and $\chi_{\rm PS}=2.2$ is the Flory interaction parameter controlling the immiscibility between protein and solvent. Equation~(\ref{Eq:E_c}) incorporates the DNA field through the local incompressibility condition
\begin{align} \label{Eq:compress}
    \phi_{\rm P}+\phi_{\rm S}+\phi_{\rm D}=1,
\end{align}
where the DNA volume fraction is represented as
\begin{align} \label{Eq:phiD}
    \phi_{\rm D}(\mathbf x)
    &= \sum_i C \exp\left(
    -\frac{|\mathbf x-\mathbf r_i|^2}{2d^2}
    \right).
\end{align}
Here, $C=0.25$, and $\mathbf r_i$ denotes the position of the $i$th discretization point along the DNA chain. Neighboring points have a stress-free separation $b\equiv|\mathbf r_{i+1}-\mathbf r_i|_{\rm no\,stress}$, with a stiff harmonic bond potential allowing small extensions while keeping bond-length fluctuations small. With the chosen parameters $C=0.25$ and $b\equiv d\equiv\lambda = 1\,\rm nm$, the maximum $\phi_{\rm D}$ for a straight DNA conformation is approximately 0.63. Steric exclusion between DNA segments is implicitly enforced by the local incompressibility constraint: overlapping DNA segments would otherwise lead to $\phi_{\rm D}>1$, violating the physical bounds on the volume fractions. For simplicity, the free-energy model in Eq.~(\ref{Eq:E_c}) assumes neutral pairwise interactions between DNA and the other two components, i.e., $\chi_{\rm PD}=\chi_{\rm SD}=0$, thereby avoiding additional interaction parameters. Even with these neutral pairwise interactions, protein preferentially wets the DNA contour because local protein enrichment near DNA reduces the energetic penalty associated with unfavorable protein--solvent contacts. This localized accumulation generates an effective capillary attraction between nearby DNA segments. We choose $\rho_0=50\,\mathrm{nm}^{-3}$ such that the resulting protein-mediated attraction remains moderate and comparable to DNA elastic energy rather than overwhelmingly dominating the system.

Because the incompressibility condition determines $\phi_{\rm S}=1-\phi_{\rm P}-\phi_{\rm D}$, the protein field $\phi_{\rm P}$ can be taken as the independent composition field. Its chemical-potential density is obtained from the variational derivative of the free-energy functional,
\begin{align}
    \frac{\mu_{\rm P}}{\rho_0k_{\rm B}T}
    &\equiv
    \frac{\delta E_{\rm c}/(\rho_0k_{\rm B}T)}{\delta \phi_{\rm P}}
    =
    \ln\frac{\phi_{\rm P}}{\phi_{\rm S}}
    + \chi_{\rm PS}
    \left[
    1-2(1+\lambda^2\nabla^2)\phi_{\rm P}
    \right]
    - \chi_{\rm PS}
    (1+\lambda^2\nabla^2)\phi_{\rm D}.
\end{align}
The spatiotemporal evolution of the protein field is governed by the Cahn--Hilliard equation,
\begin{align} \label{Eq:PDE}
    \frac{\partial \phi_{\rm P}}{\partial t}
    &= \frac{D}{\rho_0k_{\rm B}T} \nabla^2 \mu_{\rm P},
\end{align}
where $t$ is time and $D$ is the protein diffusivity. The solvent field follows directly from the incompressibility condition and is therefore not evolved independently.

Unlike the diffusible composition fields, the DNA volume-fraction field $\phi_{\rm D}$ is determined by the instantaneous conformation of the discretized chain, $\{\mathbf r_i\}$, through Eq.~(\ref{Eq:phiD}). The DNA conformation evolves according to overdamped Langevin dynamics for the discrete segment coordinates,
\begin{align} \label{Eq:ODE}
    \frac{{\rm d} \mathbf r_i}{{\rm d} t}
    &= -\frac{1}{\zeta}
    \frac{\partial E_{\rm tot}}{\partial \mathbf r_i}
    + \sqrt{\frac{2k_{\rm B}T}{\zeta}}\,
    \boldsymbol{\eta}_i,
\end{align}
where $\zeta\equiv 100\,k_{\rm B}T/D$ is the drag coefficient associated with each discretized DNA segment and $E_{\rm tot}=E_{\rm el}+E_{\rm c}$ is the total energy of the system. The deterministic force therefore includes both the DNA elastic forces and the forces arising from the dependence of the chemical interaction energy on the DNA conformation through $\phi_{\rm D}$. The second term represents thermal fluctuations, where $\boldsymbol{\eta}_i(t)$ is a three-dimensional Gaussian white-noise vector satisfying
\begin{align}
    \langle \boldsymbol{\eta}_i(t) \rangle &= \mathbf 0,\\
    \langle \boldsymbol{\eta}_i(t)\boldsymbol{\eta}_j^{\rm T}(t')\rangle
    &= \delta_{ij}\delta(t-t')\mathbf I.
\end{align}
The characteristic time scale $t_0$ used in the main text is the analytical longest Rouse relaxation time of the corresponding ideal Gaussian ring polymer \cite{doi1988theory, tsolou2010melt},
\begin{align}
    t_0 = \frac{\zeta L^2}{12\pi^2k_{\rm B}T},
\end{align}
where $L=Nb$, with $N$ being the number of
discretized DNA segments and $b=1\,\mathrm{nm}$ their spacing. This time scale
serves as a reference for normalizing the
simulation time rather than a relaxation
time measured from the simulations.

\section{Equilibrium State at Zero Supercoiling}

\renewcommand{\thefigure}{S\arabic{figure}}
\begin{figure}[!t]
\centering
\includegraphics[width=1\linewidth]{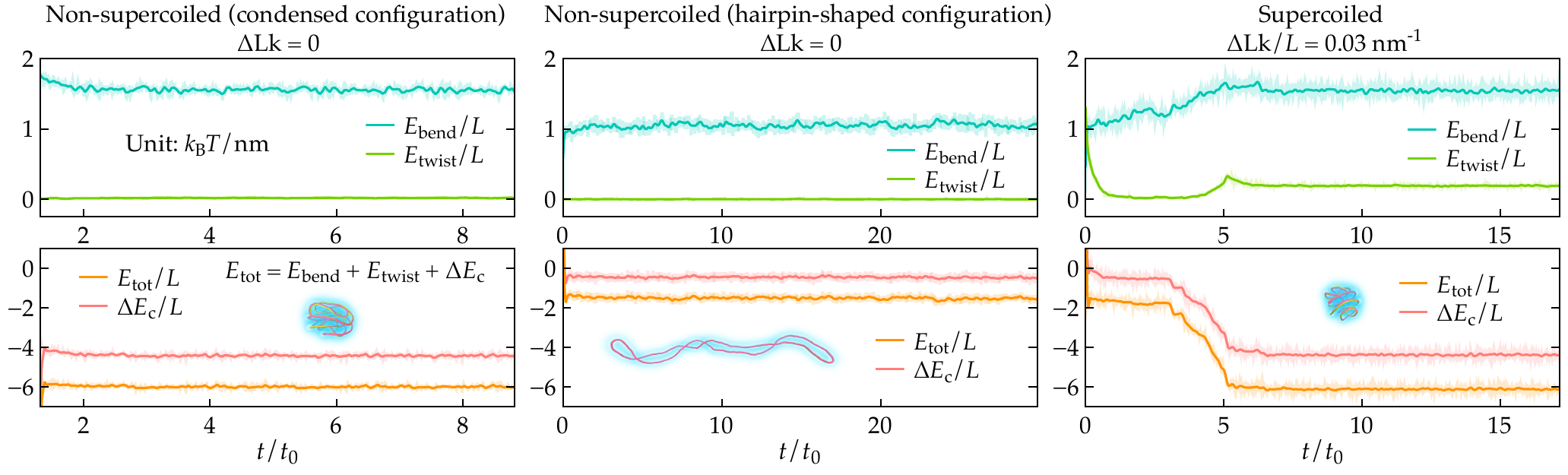}
\caption{
Time evolution of the energy contributions for different DNA topologies. The left and middle panels correspond to DNA with zero excess linking number ($\Delta{\rm Lk}=0$), while the right panel shows a supercoiled system with linking density $\Delta{\rm Lk}/L=0.03\,\mathrm{nm}^{-1}$. All simulations are initialized from a hairpin-shaped DNA contour ($L=300\,\mathrm{nm}$) composed of two parallel, end-linked segments. The middle and right panels correspond to the simulations shown in Fig.~3. For the system in the left panel, we transiently strengthen the protein-mediated capillary attraction during the initial phase ($t\lesssim1.32\,t_0$) by increasing the lattice density to $\rho_0=200\,\mathrm{nm}^{-3}$. This transient bias allows the system to overcome the nucleation barrier and drives the DNA into a condensed configuration. At $t\approx1.32\,t_0$, we restore $\rho_0$ to its baseline value of $50\,\mathrm{nm}^{-3}$ and allow the system to re-equilibrate under the original model parameters. Comparison of the left and middle panels shows that the condensed conformation has a lower total energy than the extended hairpin state, demonstrating that the latter is kinetically trapped by a substantial nucleation barrier.
}
\label{condense}
\end{figure}

In the main-text simulations (Fig.~3), non-supercoiled DNA ($\Delta{\rm Lk}=0$) does not collapse despite protein wetting, but instead remains trapped in a fluctuating hairpin configuration over the simulated timescale. However, the $\sim4.5\,k_{\rm B}T/\mathrm{nm}$ reduction in chemical interaction energy observed during the collapse of supercoiled DNA ($\Delta{\rm Lk}/L=0.03\,\mathrm{nm}^{-1}$) should in principle be sufficient to favor condensation of non-supercoiled DNA as well, since the associated mechanical penalty remains comparatively small ($\lesssim1\,k_{\rm B}T/\mathrm{nm}$). Furthermore, the analytical model in Fig.~4e predicts that, even at zero supercoiling, sufficiently compact configurations at small radius $R$ and helix angle $\gamma$ are energetically favored within the constrained plectonemic manifold. Here, we perform additional simulations to verify explicitly that a geometrically unconstrained, fully developed co-condensate is thermodynamically favored over the extended hairpin state for non-supercoiled, protein-wetted DNA.

As shown in Fig.~\ref{condense}, to access this collapsed configuration and bypass the kinetic barrier, we use the same simulation protocol as in Fig.~3 but transiently strengthen the protein-mediated capillary attraction by increasing the microscopic lattice density to $\rho_0=200\,\mathrm{nm}^{-3}$ during $t\lesssim1.32\,t_0$. This drives the non-supercoiled DNA to rapidly condense within a protein-rich droplet. After $t\approx1.32\,t_0$, we restore $\rho_0$ to its baseline value of $50\,\mathrm{nm}^{-3}$ and allow the system to re-equilibrate under the original model parameters. The resulting chemical interaction energy density $\Delta E_{\rm c}/L$ is nearly identical to that of the supercoiled condensed state. This similarity indicates that, once comparably compact conformations are reached, the chemical interaction energy is only weakly sensitive to DNA topology and is instead controlled primarily by the spatial proximity and contact between DNA segments. Because the resulting compact state has a lower total energy density than the extended hairpin configuration, we conclude that the non-supercoiled DNA in Fig.~3 is kinetically trapped, separated from the lower-energy condensed state by a substantial nucleation barrier.

\section{Phase Diagram for Hairpin Initial Configurations}

In our analytical model (Fig.~4), DNA is constrained to a uniform plectonemic geometry characterized by a radius $R$ and a helix angle $\gamma$. Exploring the energy landscape within this two-dimensional manifold yields the stability boundary plotted in Fig.~2. This boundary marks the limit of metastability at which the energy barrier separating the extended and compact local minima vanishes, allowing an extended configuration ($\gamma\approx90^\circ$) to collapse spontaneously toward a more compact state at smaller $\gamma$. While this boundary qualitatively delineates the extended and condensed regimes observed in the full simulations of Fig.~2, those simulations are initialized from a circular DNA conformation. To test the analytical prediction more directly, we therefore perform an additional set of simulations initialized from a pre-formed hairpin structure, which more closely resembles the extended branch of the constrained plectonemic pathway assumed in the reduced model.

As shown in Fig.~\ref{phase_diagram_hairpin}, the resulting DNA morphologies are largely consistent with both the phase diagram in Fig.~2 and the analytically predicted stability boundary. However, in certain parameter regimes, hairpin-initialized DNA remains kinetically trapped in an extended plectonemic state, whereas ring-initialized DNA fully condenses into a droplet. This discrepancy reflects a kinetic difference between the two initial conditions: a pre-formed linear plectoneme can face a higher barrier to complete collapse than an initially circular DNA molecule. For circular DNA, rapid conversion of stored twist into writhe generates transient branched intermediate conformations, including star-like plectonemes. By contrast, supercoiled hairpin DNA tends to form a single extended plectoneme. The branched intermediates generated from circular DNA can cooperate strongly with protein wetting, particularly near the multi-branch junctions where several DNA segments are brought into close spatial proximity, thereby facilitating the nucleation of condensation. During this rapid conformational rearrangement, protein wetting can stabilize these transient close-contact geometries, promoting capillary collapse (see Supplemental Movie~1) and helping the system bypass kinetic traps that can stall a fully developed linear plectoneme.

The behavior beyond the critical point of the reduced energy landscape requires additional interpretation. The critical point marks the termination of the bistable regime, rather than a transition to complete DNA collapse. Beyond this point, the constrained model predicts a single energy minimum at a preferred helix angle $\gamma^*$, whose value determines the degree of plectonemic compaction. As shown in Fig.~\ref{phase_diagram_hairpin}, when protein wetting is weak ($\phi_{\rm P}\lesssim0.05$), the preferred helix angle remains relatively large, and the resulting DNA conformation remains an extended plectoneme rather than collapsing into a compact droplet. Thus, the number of minima alone does not determine the extent of DNA compaction, which instead depends on the preferred plectonemic geometry. The regime emphasized in this work, however, lies within biologically relevant supercoiling densities, typically $|\sigma|\lesssim0.1$ \cite{higgins2016species, visser2022psoralen, junier2023dna}, and within the parameter range where the predicted stability boundary provides a good description of the full simulations. At more extreme supercoiling, additional structural responses of the DNA double helix, including local melting or transitions to non-canonical conformations \cite{bryant2003structural, sheinin2011underwound}, may become important and are beyond the scope of the present model.

\renewcommand{\thefigure}{S\arabic{figure}}
\begin{figure}[!t]
\centering
\includegraphics[width=0.9\linewidth]{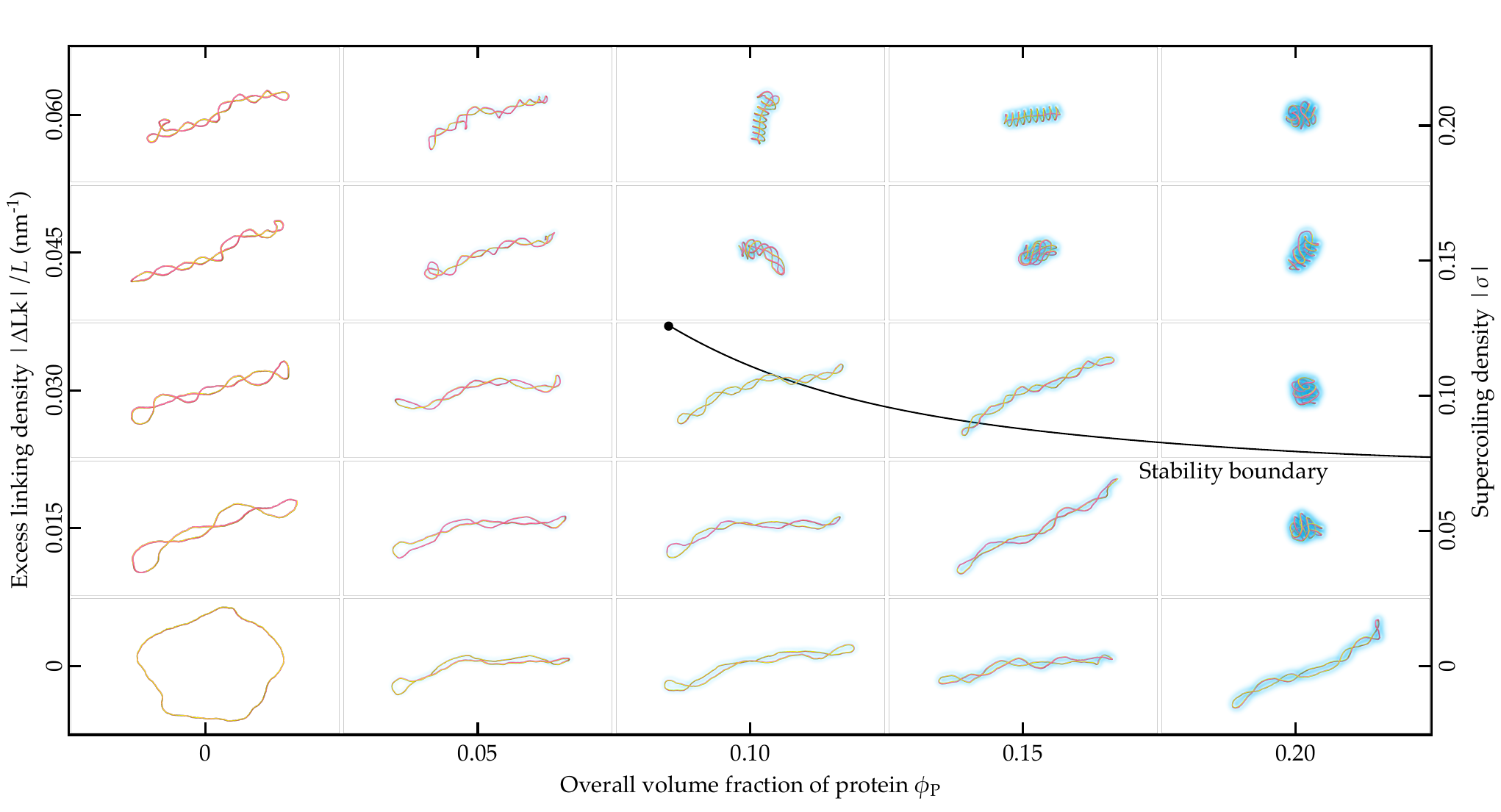}
\caption{
Phase diagram of DNA conformations as a function of the mean protein volume fraction $\phi_{\rm P}$ and excess linking density $\Delta{\rm Lk}/L$. All simulations are initialized from a closed hairpin-shaped DNA contour of length $L=300\,\mathrm{nm}$. The corresponding supercoiling density is $\sigma=h\Delta{\rm Lk}/L$, where $h\approx3.4\,\mathrm{nm}$ is the helical pitch of relaxed DNA. The stability boundary is obtained from the theory introduced in Fig.~4.
}
\label{phase_diagram_hairpin}
\end{figure}

\bibliography{Supercoiling}